\documentclass[aps,prb,showpacs,twocolumn,groupedaddress]{revtex4}

\usepackage{graphicx}
\usepackage{dcolumn}
\usepackage{amsmath}

\begin{document}

\newcommand{\mgb}{{MgB$_2$ }}
\newcommand{\be}{\begin{equation}}
\newcommand{\ee}{\end{equation}}
\newcommand{\bea}{\begin{eqnarray}}
\newcommand{\eea}{\end{eqnarray}}

\title{Spin-resolved spectra of Shiba multiplets from ${\rm Mn}$ impurities in ${\rm MgB}_2$}
\author{C\u at\u alin Pa\c scu  Moca,$^{1,2}$
 Eugene Demler,$^{3}$ Boldizs\'ar
Jank\'o,$^{4,5}$ and Gergely Zar\'and$^{1}$}
\affiliation{
$^1$ Budapest University of Technology and Economics, H-1521 Budapest, Hungary
\\
$^2$ Department of Physics, University of Oradea, Oradea, Romania
\\
$^3$ Lyman Physics Laboratory, Harvard University, Cambridge MA
\\
$^4$ Materials Science Division, Argonne
National Laboratory, 9700 South Cass Avenue, Argonne, Illinois
60439
\\
$^5$ Department of Physics, University of Notre Dame,  Notre Dame, Indiana, 46556
}
\date{\today}

\begin{abstract}
We study the effect of magnetic ${\rm Mn}$ ions on the two-band
superconductor ${\rm MgB}_2$, and compute both the total and spin
resolved scanning tunneling spectrum in the vicinity of the magnetic
impurity. We show that when the internal structure of the ${\rm Mn}$
ion's $d$-shell is taken into account, multiple Shiba states appear
in the spectrum. The presence of these multiplets could alter
significantly the overall interpretation of local tunneling spectra
for a wide range of superconducting hosts and magnetic impurities.
\end{abstract}
\maketitle
\section{Introduction}

The interaction between a single magnetic impurity and the
superconducting host reveals fundamental properties of both the
magnetic ion and the host material. This interaction was first
studied theoretically, within the framework of BCS
superconductivity. During the late sixties Shiba \cite{Shiba} showed
that a magnetic impurity pulls down from the continuum states a pair
of bound states inside the superconducting gap. Indirect indication
for the presence of finite spectral weight inside the gap of an
impure superconductor could be inferred from global probes of the
density of states. However, direct evidence for the existence of the
so-called Shiba states requires an accurate measurement of the {\em
local} density of states near the impurity. Such a measurement
became available only recently by using high vacuum, low temperature
scanning tunneling spectroscopy (STS). Yazdani and his coworkers
imaged\cite{Yazdani} the local density of states around ${\rm Mn}$
and ${\rm Gd}$ impurities deposited onto ${\rm Nb}$ single crystals.
They found clear evidence for localized states in the vicinity of
the magnetic impurities, in qualitative agreement with Shiba's
original findings, and also with their own model calculation based
on a non-selfconsistent solution to Bogoliubov-de Gennes equations.
Quantitative discrepancies, however, are also clearly present,
especially when comparing the width and spatial dependence of the
resonances to theoretical expectations. The presence of magnetic
impurity induced bound states in a superconductor was turned around
and used, both theoretically\cite{Scalapino,Flatte,Balatsky} and
experimentally\cite{Pan}, as an investigative tool to probe the
unusual ground state of the cuprate superconductors.

Although there exist some precious numerical renormalization group and 
Monte Carlo results for quantum dots attached to superconducting electrodes,\cite{belzig,egger} 
most of the theoretical studies carried out so far for magnetic impurities in
a superconductor follow Shiba's original
work, and use predominantly a classical spin model to describe the magnetic
impurity and assume a single spin one-half electron channel that
couples to the magnetic impurity.  Furthermore, the coupling is
assumed to be in the s-wave channel, and spin-orbit coupling is
generally ignored. This set of approximations worked beautifully for
most of the experiments performed so far and provided  simple,
elegant and intuitive results. However, recent advances in the
resolution, stability and processing of scanning tunneling imaging
opened the door for visualizing structures that go beyond the class
of Shiba-like models. Indeed, magnetic impurities have a more
complicated internal structure \cite{blandin}: The magnetic moments
are usually due to low-lying and crystal-field split $d$- or
$f$-levels with multiple occupancy. The aim of this paper is to
demonstrate that $(i)$ the  internal structure of the Mn impurity
has a major impact on the structure of the Shiba states, and $(ii)$
these novel features should be readily observable with the current
resolution of STS measurements. In particular, {\em multiple
channels} of charge carriers couple to the magnetic impurity through
channel-dependent coupling. The combination of these ingredients
generally leads to the appearance of {\em multiple} pairs of Shiba
states. We compute the spatial and spin structure of the scanning
tunneling microscopy (STM) spectra around the magnetic impurity and
show that these states appear as distinct resonances inside the
superconducting gap, and can be most clearly resolved in spin
resolved STM spectra.

In the following  we illustrate our results on the specific case of
${\rm Mn}$-doped ${\rm MgB}_2$, but we wish to emphasize that much
of our discussions  carry over to other systems as well
\cite{Davis}, and that our conclusions are rather general. There are
several reasons to choose the ${\rm Mn-MgB}_2$ system. Despite the
relatively recent discovery of its essentially conventional
superconducting phase, ${\rm MgB}_2$ has been thoroughly
characterized both experimentally and
theoretically\cite{PhysicaC2007}, and therefore provides an ideal
testing ground for our theoretical framework. Several materials
parameters of ${\rm MgB}_2$ are also in a convenient range for our
investigation. First, in order to observe a Shiba state by scanning
tunneling spectroscopy (STS), one needs a relatively large gap.
${\rm MgB}_2$ is a perfect candidate in this respect since it is a
conventional superconductor that has an unusually high critical
temperature\cite{Akimitsu}, $T_c = 39\, {\rm K}$. Second, ${\rm
MgB}_2$ has a hexagonal ${\rm AlB}_2$-type structure and a highly
anisotropic band structure \cite{Budko,Mazin}. As we shall see
below,  this leads to a clear separation of the multiple Shiba
states. The presence of two gaps in ${\rm MgB}_2$ has been well
established by now through a variety of spectroscopic probes
\cite{Giubileo,Szabo,JohnZ,Tsuda}. It is therefore an
interesting question, how  the presence of these two gaps influences
the structure of Shiba states. Although a series of experimental
\cite{Gonnelli,Rogacki} and theoretical \cite{Ummarino,Kortus}
investigations have been recently completed for ${\rm MgB}_2$ doped
with nonmagnetic as well as magnetic impurities, no experimental or
theoretical study has been completed for the local electronic
structure of a {\it single} magnetic impurity in this compound. This
paper now provides a detailed theoretical discussion of the single
magnetic impurity problem in ${\rm MgB}_2$ and other superconductors
where the multiple degrees of freedom of the conduction electrons
and the impurity could lead to experimentally observable
consequences.

Finally, there is another advantage for studying Shiba states in
${\rm MgB_2}$. The strong coupling, short coherence length and the
consequently robust condensate allows us to investigate the effect
of a single magnetic impurity on a superconductor in the regime
where the order parameter remains spatially constant. Here the
results of Flatt\'e and Byers \cite{Flatte} are quite valuable:
According to their calculations, for a superconductor with coherence
length $\xi k_F =10$ the relative spatial fluctuations in the local
order parameter remain below 5\%, even at the impurity site. This
result is valid for the entire range of interest $ 0 \le g \le 1$
for the dimensionless coupling $g$ (between the magnetic impurity
and the superconducting quasiparticles, see below). As shown in the
following sections of this paper, the spatially constant order
parameter provides considerable simplifications in our calculations,
and this model allows us to make experimentally testable predictions
for the presence of the multiple Shiba states in ${\rm MgB}_2$.


\section{Hamiltonian}

\subsection{Band structure calculation}

As mentioned above, ${\rm MgB}_2$ crystallizes in the hexagonal
${\rm AlB}_2$-type structure \cite{Budko} in which the $B^{-}$ ions
constitute graphite-like sheets in the form of honeycomb lattices
separated by hexagonal layers of ${\rm Mg}$ ions. Band structure
calculations \cite{Mazin} indicate that ${\rm Mg}$ is substantially
ionized, and the bands at the Fermi level derive mainly from Boron
$p$ orbitals. Four of the six $p$ bands cross the Fermi energy, and
the Fermi surface consists of quasi-$2D$ cylindrical sheets, due to
${\rm B}$ - $ p_{x,y}$ orbitals, and a $3D$ tubular network (mostly
originating from ${\rm B}$ - $p_z$ orbitals). It is believed that
both structures participate in the formation of the superconducting
state, though the gap  is very different on the tubular network and
on the cylindrical sheets.

Let us first discuss the tight-binding Hamiltonian we use and the
corresponding band structure. In spite of its simplicity, this tight
binding description is rather robust, as can be  checked by a direct
comparison to the results of more sophisticated ab-initio
 band structure and density of states (DOS) calculations.\cite{Mazin}
In the rest of the paper we shall use the following simple Hamiltonian
 to describe the normal  state of \mgb,
\begin{eqnarray}
H_0& = &\sum_{\mathbf{r} , \mathbf{r'}}\sum_{\alpha, \alpha',\;\sigma } 
\left( t_{\mathbf{r}, \mathbf{r'}}^{\alpha, \alpha'} -\mu
  \;\delta_{\mathbf{r}, \mathbf{r'}}\;\delta^{\alpha,\alpha'}\right)\times \nonumber\\
&& \left ( \Psi^{\dagger}_{ \mathbf{r}, \alpha, \sigma}
\Psi _{ \mathbf{r'}, \alpha', \sigma} + h.c. \right)
\;, \label{eq:h0}
\end{eqnarray}
where $\mu$ sets the Fermi energy and $\Psi_{\mathbf{r}, \alpha, \sigma}$ is the annihilation operator
of an electron  of spin $\sigma$ on ${\rm p}$-orbital $\alpha$ ($\alpha
=p_x,p_y,p_z$) of the  B ion  at position $\bf r$,
\begin{equation}
\mathbf{r} = \mathbf{R}+\mathbf{d}\;.  \label{eq:unit_cell_sites}
\end{equation}
The vector $\mathbf{R}$ in this expression
 points to the center of the unit cell and $\mathbf{d}$ gives the position
of the B ion within the unit cell. Note that there are two atoms per unit cell,
which shall be labeled by the index $\delta=1,2$ in what follows.
 The hopping matrix elements $t_{\mathbf{r}, \mathbf{r'}}^{\alpha, \alpha'} $
in Eq.~\eqref{eq:h0} connect  only neighboring
sites, but their value depends on the relative orientation of the p-orbitals.
Quasiparticle energies are measured from the Fermi energy, $\mu$.

\begin{figure}[h]
\centerline{\includegraphics[width=3.0in]{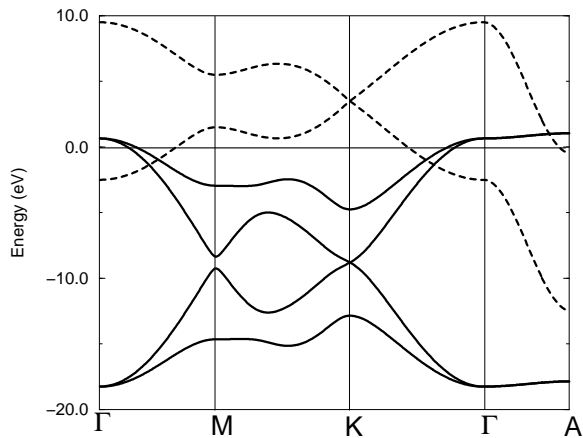}}
\vspace*{3ex} \caption{Band structure for $MgB_2$, along the
symmetry lines, computed in the framework of tight-binding model
described in Appendix A. There are six bands: four from $p_{x,y}$
orbitals (solid lines) and two from $p_z$ (dashed lines). Both $p_z$
bands and only two $p_{x,y}$ band cross the Fermi level which
corresponds to zero energy, $E_F =0$.} \label{fig:bands}
\end{figure}

The Hamiltonian above can be easily diagonalized in Fourier space. The
field operators $\Psi _{ \mathbf{r}, \alpha, \sigma}$ can be expanded
as
\begin{equation}
\Psi _{\mathbf{r},\alpha ,\sigma}  =\frac{1}{\sqrt{\Omega}}
                                                \sum\limits_{\mathbf{k}, b}
                                                 e^{i\mathbf{k\;R}}
                                     \; e_{b;\alpha, \delta }\left( \mathbf k \right )
                                      \;c_{\mathbf{k}, b, \sigma }\;, \label{eq:c}
\end{equation}
where $\Omega$ is the number of unit cells, and $ c_{\mathbf{k}, b, \sigma}$ is the annihilation operator of an
 electron in band $b$ ($b=1,\dots,6$)  with momentum $\mathbf k$, spin $\sigma$, and
 energy $\varepsilon_{\mathbf k,b}$. The band energies and the wave function amplitudes
 $e_{b;\alpha, \delta }$ are determined by the eigenvalue equation
\begin{equation}
\sum_{ \alpha', \delta'}H_{\alpha, \delta; \alpha', \delta'}\left(
\mathbf{k} \right)\; e_{b; \alpha', \delta'}=
\varepsilon_{\mathbf{k}, b}\; e_{b; \alpha, \delta}\;,
\label{eq:Hamiltonian_matrix}
\end{equation}
where $H_{\alpha, \delta; \alpha', \delta'}( \mathbf{k})$ is
essentially the Fourier transform of the hopping matrix, detailed in Appendix A.
In our tight binding model we have  six bands: Four of them derive from
$p_{xy}$ orbitals while the remaining two from $p_z$ orbitals. The
band structure obtained  is  presented in Fig. \ref{fig:bands}.
Notice that both $p_z$  ($\pi$ bands) cross the Fermi surface
but only two of the $p_{x,y}$ bands  ($\sigma$ bands), cross it.

In  the presence of superconducting order, one must modify the
Hamiltonian above and add the pairing terms,
\begin{eqnarray}
H_0 \to H_0 &=&
\sum\limits_{b,\mathbf{k},\sigma }\varepsilon _{\mathbf{k},b} \; c_{
\mathbf{k},b,\sigma }^{\dagger }c_{\mathbf{k},b,\sigma }\nonumber\\
&+ &\sum\limits_{b,\mathbf{k}}\Delta _b\left( c_{-\mathbf{k},b,\downarrow
}^{\dagger }c_{\mathbf{k},b,\uparrow }^{\dagger }+h.c.\right) .  \label{eq:H_0}
\end{eqnarray}
Here the summation goes over those  four bands that cross the Fermi energy ($b=1\dots4$).
We assume further that the superconducting gaps take only two different
values:  in the $p_{x,y}$ bands $ \Delta_{xy}\approx 7.5$meV while
for the $p_z$-bands it is   $\Delta _z\approx 2.5$ meV. In our work, we shall
neglect furthermore the position-dependence of the gaps around the magnetic impurity. This
approximation is justified by the short coherence length in \mgb, as already
explained in the introduction.

\subsection{Interaction with a magnetic impurity}

\begin{figure}[b]
\centerline{\includegraphics[width=2in]{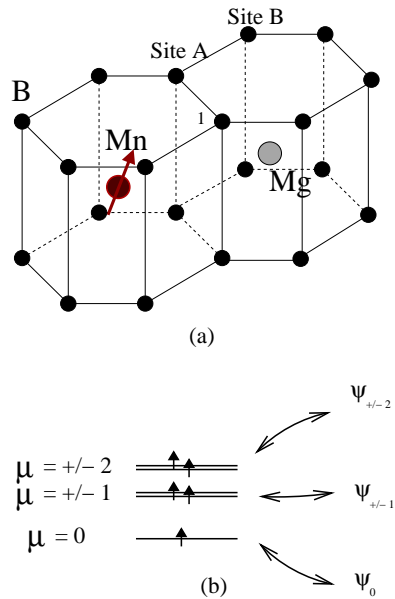}} \vspace*{3ex}
\caption{\label{fig:cage}
Fig.~a: Local environment of an ${\rm Mn}$ ion in ${\rm MgB}_2$.
The ${\rm Mn}$ ion is located in a hexagonal cage
made of ${\rm B}$ ions represented by large dots.
Fig.~b: Level structure and crystal field splitting
of the ${\rm Mn}^{2+}$ core states. Two levels
($\mu =\pm 2$ and $\mu = \pm1$) are two fold degenerate
and the level $\mu=0$ is non-degenerate. The order of the 
d-levels may depend on the details of the cristal field.}
\end{figure}

To carry out a quantitative analysis of the magnetic impurity
problem, we first need to establish how magnetic  spins couple to the
conduction band. The interaction part of the Hamiltonian
depends on the specific location and electronic structure of the
magnetic impurity considered. In what follows, we provide a
detailed analysis for ${\rm Mn}$ impurities, which have already been
doped into \mgb, though similar considerations hold for other types and positions
of magnetic impurities.
 ${\rm Mn}$ ions presumably substitute
the ${\rm Mg}$ atoms, and most likely take an ${\rm Mn}^{2+}$ configuration
with a half-filled $d$-shell and a spin $S\approx 5/2$.\cite{Mndoping}
 As shown in Fig.~\ref{fig:cage}, the five-fold degeneracy of the $d$-states is
lifted by the local hexagonal crystal field
into three multiplets that we can label by the original angular momentum
quantum numbers $\mu$  of the d-states, $|\mu\rangle$. Each of these states
is occupied by a single electron, and hybridizes through
a hybridization $V_{\mu}$ with  a specific local combination of $p$-states,
$\psi_{\mu}$ that we construct next.

The ${\rm Mn}$ ion is in the middle of a cage of 12 B ions, that we
shall label by the index $i=1,\dots,12 $. To start with, let us
first construct the local hopping Hamiltonian between the  ${\rm
Mn}$ d-orbitals and the p-orbitals of a neighboring ${\rm B}$ ion
'$i$' at position $\mathbf{r}_i$. Let us now  take a reference frame
with the ${\rm Mn}$ in the origin and the $z$-axis pointing along
the direction $\mathbf{n}_i$ of this neighboring ion. In this
reference frame, with a good approximation, only the $\tilde L_z=0$
state of the five Mn d-states hybridizes with the $\tilde L_z=0$
state of the B p-orbital. Correspondingly, the hybridization between
the impurity  and this neighbor
 can be approximated as
\be V_i =V \;\Psi ^{\dagger \|}_{{\mathbf r_i}, \sigma}  d
_{i,\sigma}^{\|} + h.c.\;, \ee where $ d _{i,\sigma}^{\|}$ is the
annihilation operator for a local ${\rm Mn}$ ${\rm d}$-orbital at
the origin oriented along the direction ${\mathbf n_i}$. Similarly,
$\Psi ^{\|} _{{\mathbf r_i},\delta _{i}, \sigma} $ is the
annihilation operator for the ${\rm p}$ state at the ${\rm B}$ site
oriented along the same direction. These operators  are related to
the operators occurring the $H_0$ by simple  rotations, \bea
\Psi^{\|}_{{\mathbf r_i},\sigma} & =&\sum _{\beta=x,y,z}
\Psi_{{\mathbf r_i},\beta, \sigma}\; n_i ^{\beta}\;,
\label{eq:operator}
\\
d_{i,\sigma}^\| &=&\sum _{\mu} \alpha_ {\mu}\left( \theta _i \right ) e^{-i
  \tilde \phi_i \mu }
d _{\mu, \sigma}\;,
\eea
where $d _{\mu, \sigma}$ refers to states with a quantization axis
perpendicular to the B planes, $\tilde\phi_i = \phi_i-\phi_1$, and
\begin{equation}
\alpha _{\mu} \left(\theta \right )= 
\left(
\begin{array}{cc}
-\frac{1}{2}\sqrt{\frac{3}{2}}\sin^2 \theta \\
-i \frac{1}{2}\sqrt{\frac{3}{2}}\sin 2\theta \\
\frac{1}{4}\left( 1+3 \cos 2\theta \right) \\
-i \frac{1}{2}\sqrt{\frac{3}{2}}\sin 2\theta \\
-\frac{1}{2}\sqrt{\frac{3}{2}}\sin^2 \theta \\
\end{array}
\right)\;.
\end{equation}

Summing over all neighboring atoms and expressing all operators
$\Psi_{{\mathbf r_i},\beta, \sigma} $
in terms of the band operators, $c_{\mathbf{k},b,\sigma}$, we then obtain the following
hybridization Hamiltonian,
\begin{equation}
H_V=\sum_{b,\mu,\sigma }V_b^{(\mu)}\left( \Psi _{b,\mu,\sigma }^{\dagger
}d_{\mu,\sigma }+d_{\mu,\sigma }^{\dagger }\Psi _{b,\mu,\sigma }\right).
\end{equation}
where the operator $\Psi _{b,\mu,\sigma }$ creates an electron with the
same local $d$-state symmetry as $|\mu\rangle$ in band $b$, and can be
expressed as
\bea
\Psi _{b,\mu, \sigma}&=&{\frac 1{\sqrt{\Omega }}}\sum_{\mathbf{k}}\widetilde{f}_{\mu, b}(
\mathbf{k})\;c_{\mathbf{k},b, \sigma }\;,  \label{eq:form}
\\
\tilde f_{\mu, b}\left( {\mathbf k}\right )&=& \sum _{i,\alpha}
\frac {\alpha _{\mu}^{*}}{A _{\mu, b}}  \left( \theta _i\right) e^{i\; \tilde \phi _i \mu}
n_i ^\alpha e_{b;\alpha,\delta _i }\left({\mathbf k} \right)
e^{i \mathbf{kR}_i}\;,\;\;\;\;
\eea
and $V_{b}^{(\mu)}=V A _{\mu, b}$. In these expressions the normalization
factor  $A _{\mu, b}$ has been determined numerically, and is defined by the
condition that  $\tilde f_{\mu,
  b}\left( {\mathbf k}\right )$ be normalized at the Fermi surface,
\begin{equation}
\frac 1{S_b}\int\limits_{S_b}d^2\mathbf{k}\widetilde{f}_{\mu, b}(\mathbf{k})
\widetilde{f}{_{\mu^{\prime}, b}^{*}(\mathbf{k})}=\delta _{\mu,\mu^{\prime }}\;.
\label{eq:orthogonal}
\end{equation}
Symmetry further implies that states belonging to the same irreducible
representation have the same hybridization: $V_b^{(\mu)}=V_b^{(-\mu)}$.

The above hybridization  Hamiltonian generates an effective exchange interaction between
the ${\rm Mn}$ spin and the conduction electrons in the ${\rm B}$ bands, since
it generates  charge fluctuations to the ${\rm Mn^{1+}}$
and ${\rm Mn^{3+}}$ states. Second order perturbation theory
in the hybridization leads  to the
effective exchange Hamiltonian:
\be
H_{\mathrm{int}}=\sum\limits_{b,b^{\prime},\mu,\alpha,\beta}
\frac 12 J_\mu^{bb^{\prime }}
\; \Psi_{b, \alpha}^{\dagger }\;\boldsymbol{\sigma}_{\alpha \beta} \cdot
\mathbf{S}\;
\Psi_{b^{\prime },\beta}
\;,  \label{eq:H_int}
\ee
where $\boldsymbol{\sigma} $ denotes the Pauli matrices,
$\mathbf S$ is the ${\rm Mn}$ spin,
and the exchange couplings are given by
\begin{equation}
J_{\mu}^{bb^{\prime }}\propto
\frac{V_b^{(\mu)}V_{b^{\prime }}^{(\mu)}}{\Delta E} = \frac{V^2}{\Delta E}
A _{\mu, b}A _{\mu, b^{\prime}}\;,\label{eq:single_coupling}
\end{equation}
with $\Delta E$ the characteristic energy of charge fluctuations.
Note that the symmetry index $\mu$ is \emph{\ conserved} in
Eq.(~\ref{eq:H_int}), thus there are five independent orbital
channels of the conduction electrons that couple to the impurity
spin. This is simple to understand on physical grounds: the
half-filled d-shell has no orbital structure. Therefore, a
conduction electron that arrives in an orbital state $\mu$  must be
scattered back to the same orbital channel. However, electrons can
be scattered between different conduction bands, it is only their
orbital label that is conserved over the scattering process.
Therefore, in the absence of superconductivity, the channel labels
play no special role, and the $S=5/2$ spin  of the Mn ion  would be
exactly screened, resulting in  a Fermi liquid state.\cite{blandin}

By construction, the exchange coupling above satisfy
$J_{\mu}^{bb^{\prime }} =\sqrt{J_{\mu}^{bb}J_{\mu}^{b^{\prime
}b^{\prime }}}$, and furthermore, they are equal  in channels $\pm
\mu$ by symmetry. From Eq. (\ref{eq:single_coupling}) it also
follows that all the results depend only on a single dimensionless
coupling proportional to $V^2/\Delta E$. We define this coupling as
\begin{equation}
g \equiv
\frac 15 \sum_{\mu,b} \varrho_{b} J_{\mu}^{b b}\;,
\end{equation}
with $\varrho_b$ the  density of states
 at the Fermi energy in band ${\rm b}$ for one spin direction.
Furthermore, in the rest of this  paper
we shall  only consider the  classical limit,
$S\rightarrow \infty $ with $J_m^{bb^{\prime }}S=finite$. In this limit
 the impurity has no dynamics and we can solve the problem exactly.

To close this subsection let us introduce
 Nambu spinors, $\Phi _{\mathbf{k},b}= \{\Phi _{\mathbf{k},b}^{\tau\sigma} \}$, \cite{Nambu}
\begin{equation}
\Phi _{\mathbf{k},b}\equiv \left (
\begin{array}{cc}
c_{\mathbf{k},b,\uparrow }\\
c_{\mathbf{\ k},b,\downarrow } \\
-c_{-\mathbf{k},b,\downarrow}^{\dagger}\\
c_{-\mathbf{k} ,b,\uparrow }^{\dagger}
\end{array}
\right )\;.
\end{equation}
The introduction of these spinors shall simplify our calculation
considerably in the following sections. We can  rewrite the
Hamiltonian in terms of these in a compact form,
\begin{eqnarray}
H &=&\sum_{\mathbf{k},b}\Phi _{\mathbf{k},b}^{\dagger }(\hat {\varepsilon} _{
\mathbf{k},b}\tau ^z+\hat{\Delta} _b\tau ^x)\Phi _{\mathbf{k},b}  \label{eq:nambu}
\\
&+&\sum\limits_{\mathbf{k},\mathbf{k}^{\prime},b,b^{\prime },\mu}\frac
12J_{\mu}^{bb^{\prime }}\;\widetilde{f}_{\mu,b}^{*}(\mathbf{k})\Phi _{\mathbf{k}
,b}^{\dagger }\;\boldsymbol{\sigma }\cdot \mathbf{S}\;\Phi _{\mathbf{k} ^{\prime
},b^{\prime }}\widetilde{f}_{\mu,b^{\prime }}(\mathbf{k}^{\prime })\;,\;\;\;
\nonumber
\end{eqnarray}
where the $\tau ^i$'s denote Pauli matrices acting in the pseudospin
(charge) index of the Nambu spinor. In course of the derivation we
made use of time reversal symmetry that implies $ \widetilde
f_{-\mu, b}\left( -{\mathbf k} \right) =\left(
-1\right)^{\mu}\widetilde f_{\mu, b}^{*}\left({\mathbf k } \right)
$, and doubled the Hilbert space so that the components of the Nambu
spinors in Eq.\eqref{eq:nambu} must be considered as independent
variables.

\section{Green's function formalism}
In this section we shall discuss how the above Hamiltonian can be treated
within the Green's function formalism.
In the classical limit the interaction with the impurity in Eq.~\eqref{eq:nambu}
reduces to a spin-dependent potential scattering and, as we show below in detail,
the problem can be solved exactly.

In the non-interacting case, $J_{\mu}^{bb^{\prime }}=0$, 
Green's function 
is given by:
\begin{equation}
G^{(0)}_{b}(\mathbf{k},\omega )=
    \frac{1}{ {\rm i}\omega -\hat{\varepsilon} _{\mathbf{k},b}\tau ^z-
    \hat {\Delta} \tau ^x}\label{eq:free_green_function}
\end{equation}
and it is a $16\times 16$ matrix, diagonal in the band indices.
In this expression $\hat{\varepsilon} _{\mathbf{k},b} $
and $\hat {\Delta}$ are also diagonal in band indices.

\begin{figure}[b]
\centerline{\includegraphics[width=7cm]{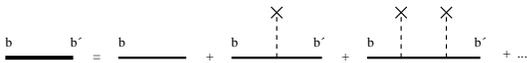}} \vspace*{3ex}
\caption{Diagramatic expansion for the Green's function when
multiple scattering on the impurity site is considered.
The solid line represents the full Green's function, the thick
line represents the non-interacting part of the Green function.
Each cross represents a scattering on an impurity site, and the
dotted line stays for the impurity scattering potential.
$b$ and $b'$ stay for band indices.
}
\label{fig:diagram}
\end{figure}
In the presence of impurity scattering we can treat the scattering
perturbatively, and use  multiple scattering theory to sum up the
series to all orders. The diagrammatic expansion of the Green's
function is represented in Fig.~\ref{fig:diagram}. In the first
order of perturbation theory the self-energy is given by
\begin{equation}
\Sigma^{(1)}\left(\mathbf{k, k'}, \omega \right)=
               \sum_{\mu}\widetilde f_{\mu}^{*}\left( \mathbf k \right)
               \;\frac{1}{2}J_{\mu}\left( \boldsymbol{\sigma}\mathbf{S}\right )
               \;\widetilde f_{\mu}\left( \mathbf k' \right)
\end{equation}
and is independent of the energy $\omega$. Here we deliberately separated the
form factors from the rest of the expression. The next order contribution gives
\bea
\Sigma^{(2)}_{bb'}\left(\mathbf{k, k'}, \omega \right) &= & 
\sum_{\mu, \mu', b'',\mathbf q}
\hat {C}^{\mu}_{bb''}(\mathbf k ,\mathbf q)
\;\left( \boldsymbol{\sigma}\mathbf{S} \right )\times\\ 
&& G^{(0)}_{b''}(\mathbf{q},\omega )
\;\left(\boldsymbol{\sigma}\mathbf{S}\right)\;
\hat {C}^{\mu'}_{b''b'}(\mathbf q ,\mathbf k')\; ,
\nonumber
\eea
where we introduced the notation, $\hat{C}^{\mu}_{bb'}(\mathbf k ,\mathbf q) =
1/2 \widetilde f_{\mu,b}^{*}\left( \mathbf k \right)
J_{\mu}^{bb'}
\widetilde f_{\mu,b'}\left( \mathbf q \right) $.
After summing over the momentum (as explained in Appendix B) and using the orthogonality
of the form factors $\widetilde f$'s at the Fermi surface we end up with
the following expression,
\bea
&&
\Sigma^{(2)}_{bb'}\left(\mathbf{k, k'}, \omega \right) = 
\\
&& \phantom{nn}= \sum_{\mu}\widetilde f_{\mu,b}^{*}\left( \mathbf k \right)
               \left[\frac{1}{2}\hat{J}_{\mu}\mathbf{S}
               \hat{F}\left(\omega \right)
               \frac{1}{2}\hat{J}_{\mu}\mathbf{S}\right]_{bb'}
               \widetilde f_{\mu,b'}\left( \mathbf k' \right)\;,\nonumber
\eea
where $\hat{F}\left(\omega \right)$ denotes the matrix,
\be
 F_{bb'}(\omega )=\delta_{bb'}\varrho_b\int\limits_{-D}^D d\varepsilon \;{\frac
1{\omega -\varepsilon \tau ^z-\Delta _b\tau ^x}}\;,
\ee
with $D$ a high-energy cut-off that can be removed in the end of the calculation.

Higher order terms can be handled in a similar way.
The final expression for the
Green's function is simply:
\begin{eqnarray}
&&G_{bb'}(\mathbf{k},\mathbf{k}^{\prime },\omega )=\delta _{
\mathbf{k,k^{\prime }}} \delta_{bb'} G^{(0)}_{b}(\mathbf{k},\omega ) \label{eq:full}
+G^{(0)}_{b}(\mathbf{k},\omega )\times\nonumber\\
&&\sum\limits_{\mu}{\frac 1\Omega }\;\widetilde{f}
_{\mu,b}^{*}(\mathbf{k})\left[\hat{T}_{\mu}(\omega )\right]_{bb'}\;\widetilde{f}
_{\mu,b'}(\mathbf{k}^{\prime })\;G^{(0)}_{b'}(\mathbf{k}
^{\prime },\omega )\;.  
\end{eqnarray}
By the orthogonality relation Eq.~\ref{eq:orthogonal}, the quantum number $\mu$
is conserved. Therefore the T-matrix $\hat{T}_{(\mu)}$ can be computed independently
for each channel $\mu$ and is given by the following expression:
\begin{equation}
\hat{T}_{\mu}(\omega )=\hat{J}_{\mu}\mathbf{S\;\sigma }/2\;\left[ 1-\hat{F}
(\omega )\hat{J}_{\mu}\mathbf{S\;\sigma }/2\right] ^{-1}\; ,  \label{eq:T}
\end{equation}
where $\hat{F}(\omega)$ denotes the diagonal matrix
$\hat{F}(\omega)=\hat{F}_{bb'}\delta_{bb'}$.
Note that $\hat{F}\left(\omega \right)$ is diagonal in the spin
labels. Therefore,  even order terms in the T-matrix are spin independent.
These terms can therefore be referred to as the ``charge scattering
channel''.
Odd order terms, on the other hand, give
spin-dependent contributions and can be referred to as a ``spin channel''.
The even (charge) channel can be directly resolved using STM technique while
for experimental observation of the odd (spin) channel
contributions spin-resolved-STM is needed.

Impurity bound states and resonances can be identified from the pole
structure of the T-matrices: True bound states correspond to zeros of the
determinants $\mathrm{det}\{\hat{T}_{\mu}^{-1}(\omega )\}$ on the real axis,
and must satisfy $|\omega |<\Delta _b$ for all bands. Zeros in the vicinity
of the real axis, on the other hand, correspond to resonances. We found that
each channel generates a bound state, but two of them are doubly degenerate by
symmetry ($\mu\to-\mu$).

It is, in general, impossible to find the poles of the $\hat{T}$
matrix analytically, and numerical calculations are needed. However,
it is generally accepted that, at least phenomenologically,
superconductivity in ${\rm MgB_2}$ can be explained using a two-band
model. With this simple assumption, the positions of the resonances
are given by the following equation:
\begin{eqnarray}
\ &(1-g_{\mu}^{11}\alpha _1(\pm E))(1-g_{\mu}^{22}\alpha _2(\pm E))=&  \nonumber \\
&\ (g_{\mu}^{12})^2\alpha _1(\pm E)\alpha _2(\pm E)\;,  \label{eq:implicit}
\end{eqnarray}
where $g_{\mu}^{bb^{\prime }}\equiv \pi S\sqrt{\varrho _b\varrho _b}
J_{\mu}^{bb^{\prime }}/2$ denote the dimensionless couplings in channel $\mu$, and
\[
\alpha _b(\omega )=\left( {\frac{\Delta _b+\omega }{\Delta _b-\omega }}
\right) ^{1/2}\;.
\]
In the present case these equations further simplify due to the relation $
g_{\mu}^{11}g_{\mu}^{22}=(g_{\mu}^{12})^2$ to
\begin{equation}
g_{\mu}^{11}\alpha _1(\pm E)+g_{\mu}^{22}\alpha _2(\pm E)=1\;.
\end{equation}
In the limiting case of $J_{\mu}^{12}=0$, Eq.~\eqref{eq:implicit} would give rise
to two pairs of Shiba states \cite{Shiba} for each channel $\mu$
corresponding to two independent bands.
Exchange coupling between the two bands, however, removes half of these
resonances.
Similarly, in the realistic situation we
 thus obtain \emph{five pairs} of Shiba states corresponding to the five
channels, but two pairs of them are two-fold degenerate because of the symmetry $
J_{\mu}^{bb^{\prime }}=J_{-\mu}^{bb^{\prime }}$.

\section{Density of States}

Our main purpose is to compute the local tunneling density of states
(LDOS) and the spin resolved density of states near a magnetic
impurity for various geometries. To obtain a quantitative estimate
for the STM spectra we performed a lengthy, but straightforward
tight-binding calculation to determine numerically the form factors
$\widetilde {f}_{\mu,b}({\mathbf k})$, the exchange couplings and
the electronic wave functions above in various geometries.

The differential conductivity $dI/dV$ measured by STM is proportional
with the local density of states which can be calculated as the imaginary part of the retarded
position dependent local Green's function:
\bea
&&\varrho_{c,\alpha}({\bf r},\omega) = - {1\over 2\pi} {\rm Im} {\rm Tr}
\left \{ G({\bf r},p_\alpha, \omega) {1 + \tau_z \over 2} \right
\}\;,
\\
&&G({\bf r},p_\alpha, \omega) =
\nonumber
\\
&&\phantom{nn}= \frac1 \Omega \sum_{\mathbf{k},\mathbf{k'}, b,b'}
 {e^{-i(\mathbf{k}-\mathbf{k'})\mathbf{R}}} \; e^{\ast}_{b,\alpha\;\delta}
e_{b,\alpha\;\delta} G_{b,b'}(\mathbf{k},\mathbf{k'},\omega)\;.
\nonumber
\eea
Similar to the charge density of states, we can also define the spin density
of states as
\be
\varrho_{s,\alpha}({\bf r},\omega) = - {1\over 2\pi} {\rm  Im} {\rm Tr}
\left \{ G({\bf r},p_\alpha, \omega) \;{\boldsymbol \sigma }
{\mathbf n}\; {1 + \tau_z \over 2} \right \}\;,
\ee
where  $\mathbf n$ is a unit vector pointing in the direction along
which we measure the spin density of states.

\begin{figure}[h]
\centerline{\includegraphics[width=3.0in]{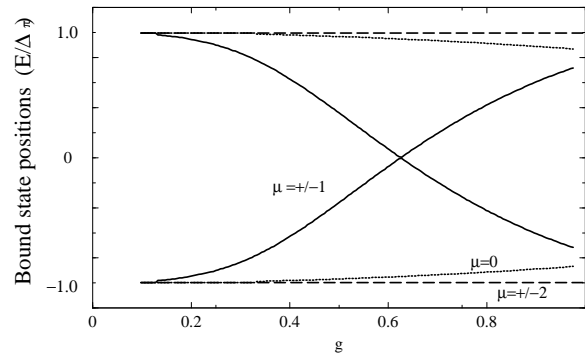}} \vspace*{3ex}
\caption{Position of the quasiparticle poles for separate channels ${\mu}$. The 
resonances corresponding to the $\mu=\pm 1$ channels are well resolved for any 
value of the coupling strength $g$ (solid lines). 
For large enough $g$, ($g> 0.6$) the 
$\mu=0$ resonance moves inside the gap (dotted lines). 
The couplings corresponding
to the $\mu =\pm 2$ channels are much smaller that those in the other channels
so the resonances for $\mu =\pm 2$ channels are still merged with 
the superconducting peaks at energy $\Delta_{\pi}$ (dashed lines).}
\label{fig:resonance_position}
\end{figure}
The equations above refer to the case where the impurity is embedded in the
bulk. However, both $\varrho_{s,\alpha}$ and $\varrho_{s,\alpha}$ can be
computed  easily from the analogue of Eq.~(\ref{eq:full}) for other boundary
conditions too, once  the wave functions  appearing in Eq.~\eqref{eq:c} are known.
In the following subsections, we first compute the LDOS for an impurity in the
 bulk. Then  we study the effect of a semi-infinite half-plane
with the ${\rm Mn}$ impurity above and below the first ${\rm B}$ layer.

\subsection{Impurity in the bulk}

\begin{figure}[h]
\includegraphics[width=7cm]{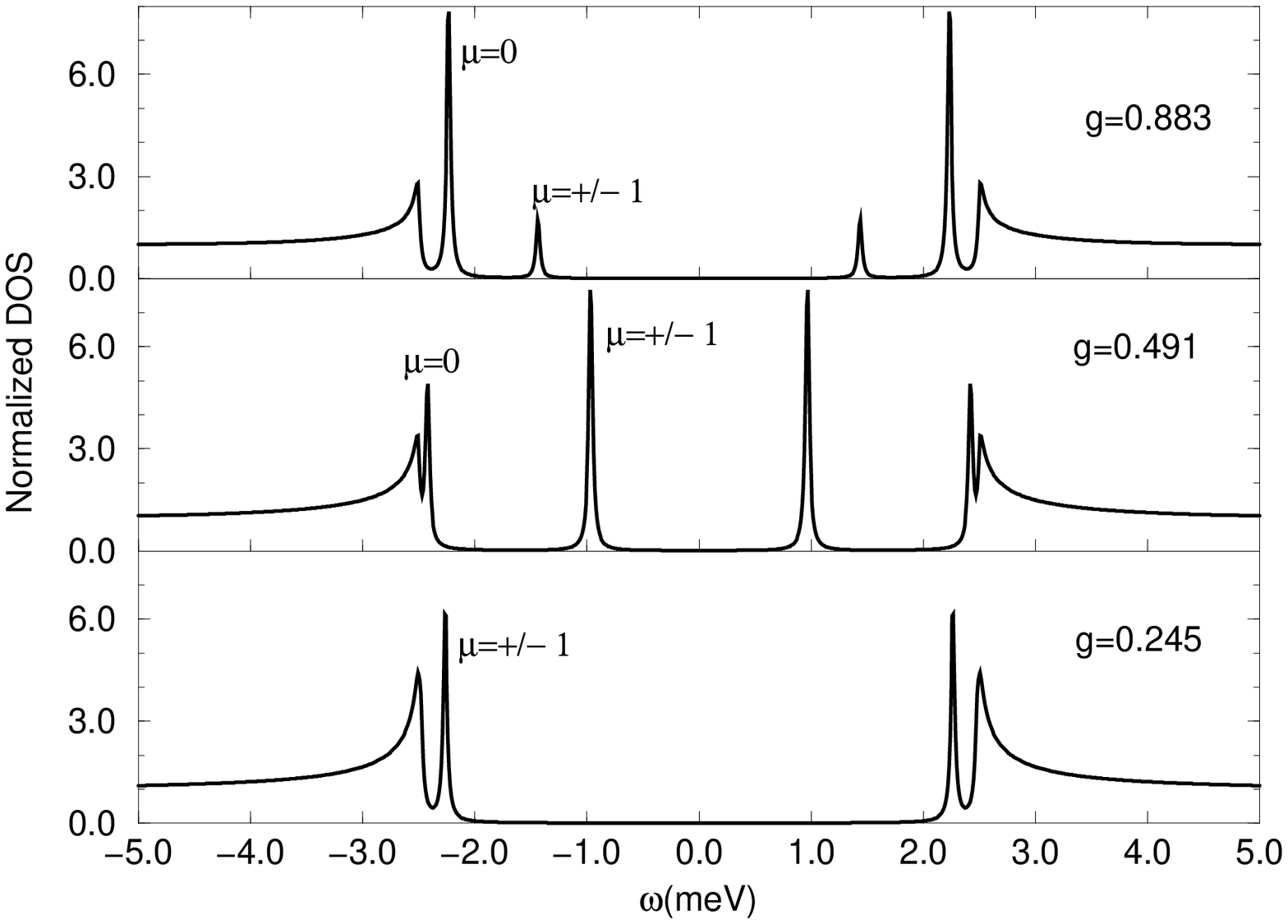}
\includegraphics[width=7cm]{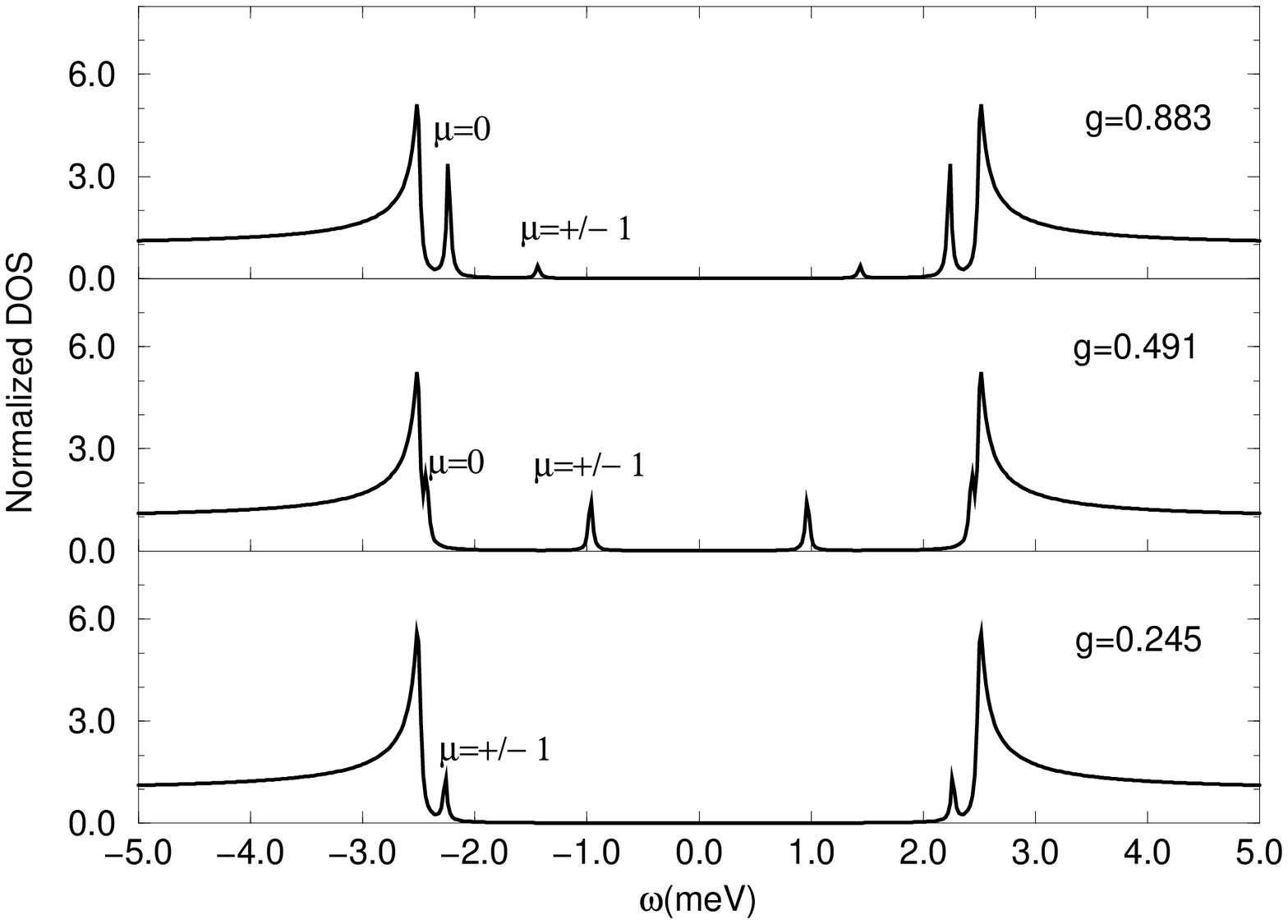}
\caption{ Upper panel: Normalized LDOS at the nearest-neighbors Boron sites
(labeled A in Fig. \ref{fig:cage}) for the $p_z$-orbitals,
for different coupling constants. Lower panel: Normalized LDOS
at the next-nearest neighbors Boron sites
labeled B in Fig. \ref{fig:cage}}
\label{fig:DOS_A_bulk}
\end{figure}

As a first step, we  identify the positions of the resonances for each channel separately from
the poles of the $\hat T$ matrix. In Fig. \ref{fig:resonance_position}
we show the positions of the bound states and resonances obtained
as a function of the dimensionless coupling $g$.
The corresponding normalized $p_z$ LDOS at the  B sites next to the Mn impurity
is  presented in  Fig. \ref{fig:DOS_A_bulk} for different values of  $g$.
Due
to hexagonal symmetry all B sites around the magnetic impurity have
the same LDOS.  For small values of $g$ the bands
are slightly interacting and only the most strongly coupled
$\mu =\pm 1$ channels give rise to a well resolved resonances in the gap
for $g\le 0.4$. Increasing the
coupling $g$, the bands are more strongly interacting and
the resonances corresponding to $\mu =0$ channel move
inside the gap too. This is accompanied, on one hand, by a transfer of
weight between resonances and secondly by a shift in
position of each resonance. We also observed
small features at
energies $\omega=7.5 {\rm meV}$, i.e. at the energy corresponding to
$\Delta_{\sigma}$, due to the coupling between the bands (not shown in this figure).

Fig. \ref{fig:DOS_A_bulk} also shows the   density of states at the
next-nearest-neighbor sites. The wave functions of the Shiba states
and thus the amplitudes of the  corresponding resonances in the
spectrum depend a lot on the tunneling position: The weight and the
amplitude of the resonances decreases considerably while their
position remains unchanged. This suppression reflects the local
structure of Shiba states. At the same time, the coherence peaks
near the superconducting gap edge gain some spectral weight, but
they are still quite reduced compared to the bulk. Further away from
the impurity site the superconducting coherence peaks are completely
restored and the bound states have negligible amplitudes. For
generic values of the exchange coupling usually two well-separated
pairs of resonances can be observed, corresponding to the $\mu=\pm1$
and $\mu=0$ channels. The exchange couplings in channels $\mu=\pm 2$
are much smaller than those in channels  $\mu=\pm 1$ and $\mu=0$,
and therefore the corresponding bound state are merged with the
superconducting coherence peak.

\begin{figure}[h]
\centerline{\includegraphics[width=3.0in]{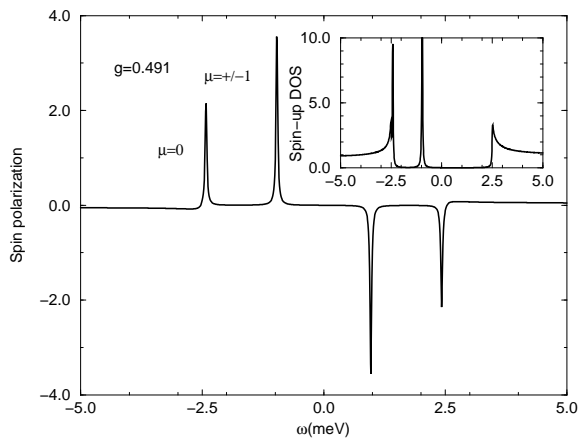}} \vspace*{3ex}
\caption{Spin polarization (odd part of the spectrum
and spin-up density of states for Site A in Fig. \ref{fig:cage}}
\label{fig:DOS_A_bulk_polarization}
\end{figure}

The Shiba states are also strongly spin-polarized, as is obvious
from the spin polarization in the local density of states shown in
Fig.~\ref{fig:DOS_A_bulk_polarization}. This fact has an important
consequence from the point of view of the observability of these
bound states. As always, a sharp local spectroscopic feature could
be difficult to detect if it is overshadowed by the intense
continuous background of the superconductor. However, the background
continuum in a superconductor is not, generally speaking, spin
polarized. Thus, even if a Shiba peak happens to be close to one of
the otherwise dominant BCS coherence peaks, a spin-polarized STM can
distinguish the Shiba states from the continuum,\cite{spinpol} since
the asymmetric part of the spin-polarized spectrum has sharp peaks
at the resonances but is predicted to be featureless otherwise.
Therefore spin-polarized STM is clearly an ideal tool to identify
the multiple Shiba states.

\subsection{Impurity in the vicinity of a surface}

As we mentioned already, the effect of a surface
can be taken into account
by simply modifying the
wave functions that appear in the expansion of the operators
$\Psi_{\mathbf{r},\alpha ,\sigma}$,
\be
\Psi _{\mathbf{r},\alpha,\;\sigma}  =       \sum\limits_{\mathbf{k_{\bot}}, k_z , b}
\varphi_{\mathbf{k_{\bot}}, k_z , b}(\mathbf{r})\;  c_{\mathbf{k_{\bot}},k_z, b, \sigma } \;.\label{eq:c1}
\ee
Here $\mathbf{k_{\bot}}$ is the in-plane momentum and
   $k_z$ is the momentum perpendicular to the surface. Note that
the surface breaks translational symmetry along the $\hat z$
direction, and therefore only $k_z>0$ values are permitted. The wave
functions above must satisfy the appropriate boundary conditions,
and can be expressed within our tight binding formalism as
\begin{equation}
\varphi_{\mathbf{k_{\bot}}, k_z , b}(\mathbf{r}) =  e_{b;\alpha, \delta }\left( \mathbf{k_{\bot}}, k_z \right )\;
                                                 e^{i\mathbf{k_{\bot}R_{\bot}}}
                                      \sqrt{2}\; \sin\left( k_z Z  \right)\;,                                     
\end{equation}
with  $Z=0$ corresponding to the first layer in the vacuum.

\begin{figure}[h]
\includegraphics[width=3.0in]{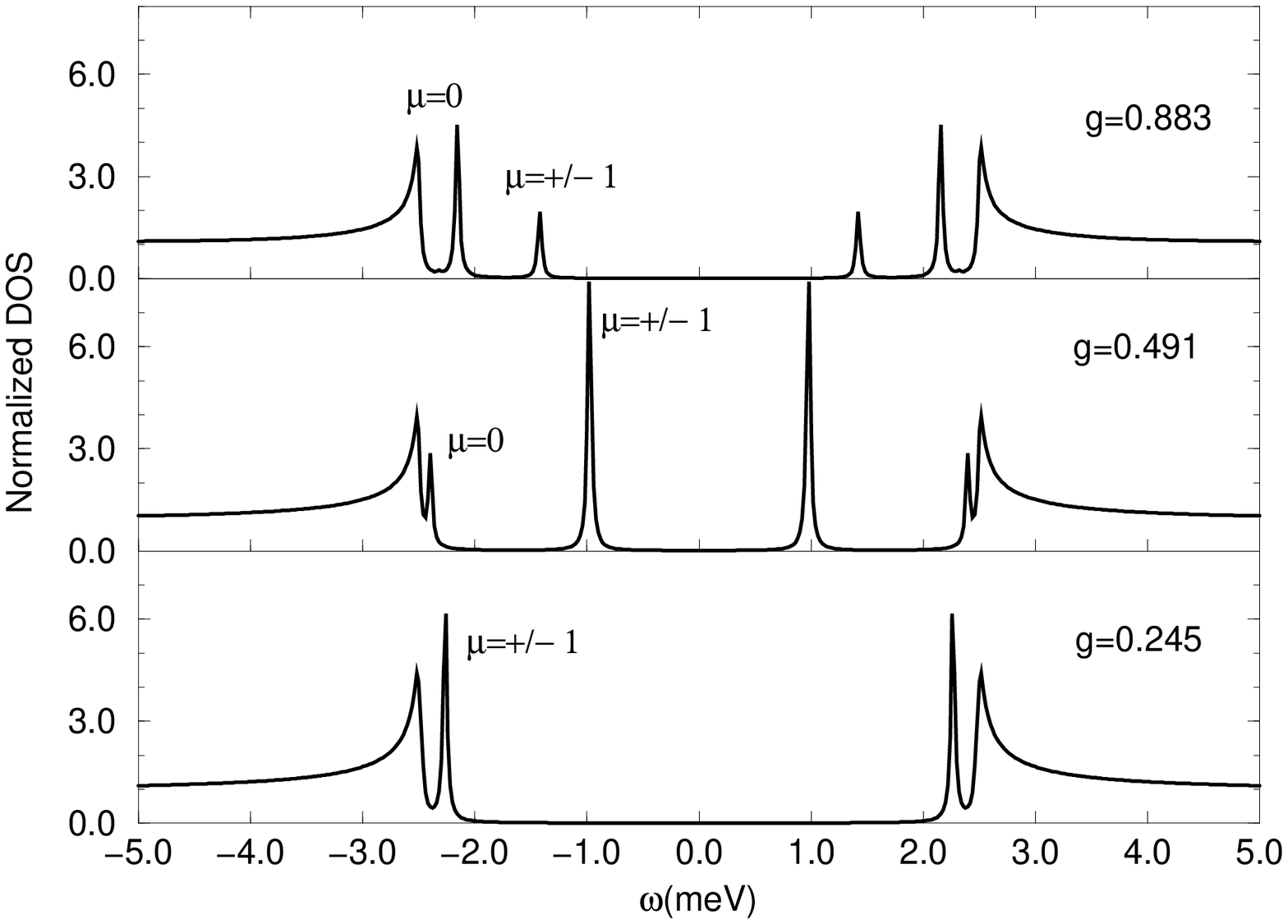}
\includegraphics[width=3.0in]{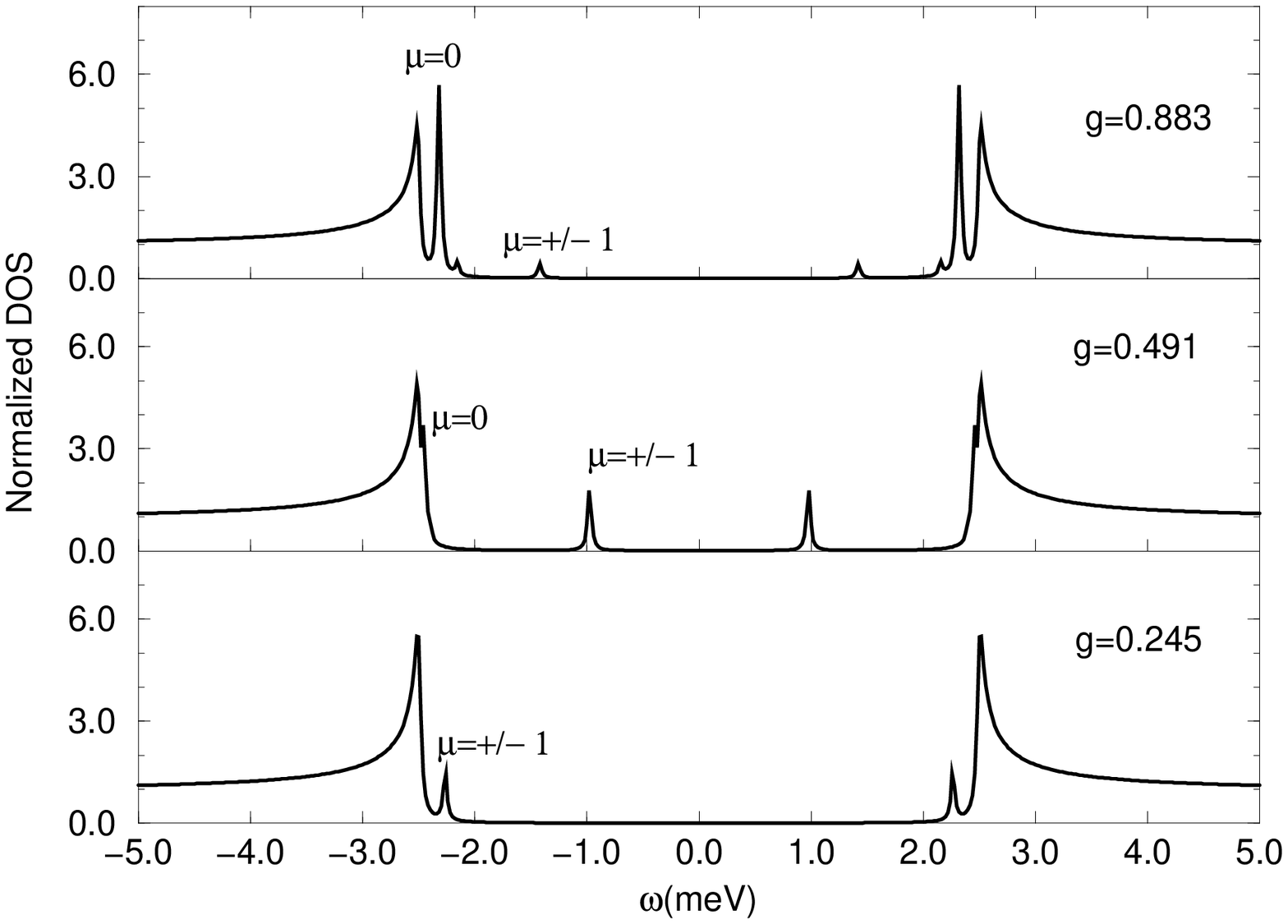}
\caption{ Upper panel: Normalized density of states for
site A  as function of frequency for different values of $g$ in the
geometry when the ${\rm Mn}$  impurity is below the first ${\rm B}$ layer.
Lower panel: Normalized LDOS at the next-nearest
neighbors Boron sites B of Fig. \ref{fig:cage}, when the $Mn$
impurity is just below the first layer}
\label{fig:DOS_A_inside}
\end{figure}
Our calculatuions for an impurity in the bulk can easily 
be extended to this case as well with minor modifications.
If the magnetic impurity is well inside the bulk, we recover the results
 discussed in the previous subsection.
In Fig. \ref{fig:DOS_A_inside} (upper panel) we represent
the LDOS at nearest neighbor B atoms for the case when the Mn impurity is
below the top B layer. The amplitudes of the resonances are slightly reduced in
this case compared to the bulk
system  and also the positions are modified due to the
local density of states that is slightly modified in the vicinity of the surface.
Moving
away from the impurity, the weights of the resonances  start to decrease
and the superconducting coherence peaks are gradually recovered.
In this configuration, at sites more than two lattice constants
away from the impurity site the superconducting coherence peak is already completely
recovered.

\begin{figure}[h]
\centerline{\includegraphics[width=3.0in]{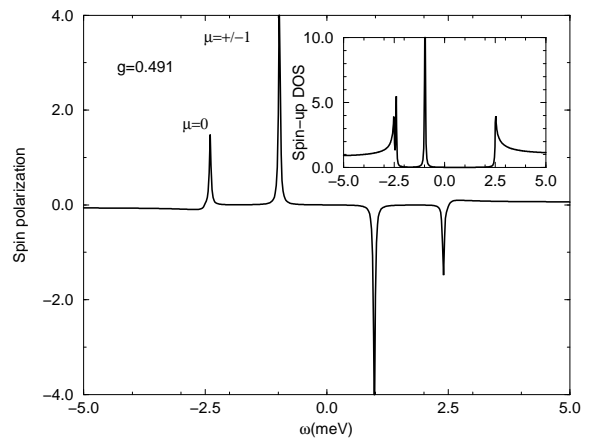}}
\vspace*{3ex} \caption{Spin polarization and spin resolved density
of states for Site A of Fig. \ref{fig:cage}}
\label{fig:DOS_A_inside_polarization}
\end{figure}
For spin-resolved scanning tunneling spectroscopy, the
tunneling current can be separated into an unpolarized part $I_0$,
which depends only on the LDOS, and a spin-polarized contribution
$I_p$ given by the projection of the local
magnetization density at the tunneling site onto the magnetization direction
of the tip. The spin-polarized contribution to the local differential
conductivity is therefore proportional to the magnetization
density, $dI_p/dV\propto P_T \cos{\theta} \varrho_s (\mathbf{r}_i,\omega=eV)$,
where $P_T$ denotes the polarization of
the tip, $\theta$ is the angle between the magnetization axes of
the tip and the impurity spin.

In Fig. \ref{fig:DOS_A_inside_polarization} we present the  the
local spin polarization at site A for  $g =0.491$. For the same
reasons as before, only  the contribution of the $p_z$ orbitals is
shown. The relative orientation of the impurity spin and the tip can
also be fixed by a small external magnetic field in these
experiments. However, the angle $\theta$ is not arbitrary even in
the absence of an external field, since in the vicinity of a
ferromagnetic  STM tip a magnetic impurity would be presumably
aligned with the magnetization of the tip due to stray fields. The
most important feature we observe is a transfer of weight from
states in the gap to states in the continuum due to the inter-band
coupling through the magnetic impurity. The inset presents the total
spin-polarized tunneling density of states for the same coupling
$g$, for a complete polarization of the tip, $P_T =1$ and a perfect
alignment, $\theta=0$.

\section{Conclusions}

We presented a detailed theoretical investigation of the effect of a
single ${\rm Mn}$ magnetic impurity on the superconducting
properties of ${\rm MgB_2}$. Our description is based on a
microscopic model which assumes nearest neighbors hopping from the
localized orbital of the ${\rm Mn}$ to the neighboring ${\rm B}$
orbitals. We have shown that a magnetic impurity generally induces
multiple Shiba states in the electronic structure of ${\rm MgB}_2$.
In particular, for ${\rm Mn}$ we found five pairs of Shiba states in
the gap, two of which were two-fold degenerate. We have taken into
account realistic band structure and the effect of surface states on
the local spectrum. Our calculation of  both conventional and
spin-resolved STM \cite{spinpol} spectra near the impurity site
showed that these states can be clearly resolved by both methods.
Similar multiple Shiba states should appear in other superconductors
due to the internal structure of the magnetic impurity.

It is intriguing to speculate what these local probes will
eventually see in an actual experiment. Clearly, despite decades of
pioneering investigation, local spectroscopy of spin impurity states
in a superconductor still has the potential of revealing new
features that have not yet been documented. For example, our
calculations assume classical spin degrees of freedom, whereas the
experimental measurements could reveal - besides a classical
behavior - effects of screening of a quantum spin by the
superconductor, leading to either a full screening or a reduction of
the effective spin carried by the impurity.  The quantitative
discussion of such effects goes beyond the scope of the present paper.
Nevertheless, our calculations will provide an important benchmark
for comparison with experiments, a benchmark that includes, for the
first time, the presence of multiple channels of scattering.

The results we obtained in this paper are relevant and relatively
easy to generalize for other compounds. For example, recent STM
measurements have focused on ${\rm Ti}$ impurities in another multiband
superconductor ${\rm Sr_2RuO_4}$. While these experimental results are
preliminary as the magnetization state of ${\rm Ti}$ is not clear, and
there are several differences between ${\rm MgB}_2$ and ${\rm Sr_2RuO_4}$,
it is clear that our framework provides a suitable platform for
studying ${\rm Sr_2RuO_4}$ as well. As we mentioned above,
${\rm MgB}_2$ crystallizes in the hexagonal ${\rm AlB}_2$-type
structure,\cite{Budko} and the band structure of ${\rm MgB}_2$ is also
somewhat peculiar. Nevertheless, as shown in the seminal paper of
Nozi\`eres and Blandin \cite{blandin},  
although the form of the
exchange Hamiltonian depends a lot on the specific material and point group
considered, in most  cases, similar to ${\rm Mn}$-doped ${\rm MgB}_2$, several channels
of conduction electrons couple to the local impurity degrees of
freedom, and result in multiple Shiba states. Therefore, 
that the appearance of multiple Shiba states is a rather general
phenomenon.

An interesting result of our analysis is that, although it may be difficult to
resolve a Shiba state close to the coherence peak with conventional STM
methods, the antisymmetrical part of a spin-resolved STM clearly separates 
these states in the STM spectrum. The weight of a given pair of Shiba states may,
however, be very sensitive to the particular atomic state into which electrons 
tunnel from the STM tip, and depends also on  the precise position of the tip.

\begin{acknowledgments}
We are grateful to D. Agterberg, G. Crabtree, J.C. Seamus
Davis, M. Iavarone, G. Karapetrov, I. Mazin, K. Tanaka,
A. Yazdani, and J. Zasadzinski for useful discussions. This work
was supported by the U.S. Dept. of Energy, Office of Science, under
Contract No.~W-31-109-ENG-38, Hungarian Grants No. OTKA NF061726, T046267,  K73361, and
Romanian Grant No. CNCSIS 1/780/2007.
B.J. was also supported by NSF-NIRT awards DMR02-10519 and ECS-0609249, 
and the Alfred P. Sloan Foundation.
\end{acknowledgments}

\appendix
\section{ Tight-binding Hamiltonian}

In this Appendix we present the basic results obtained
from  the tight-binding analysis of the bulk system.
The matrix elements of the Hamiltonian given
in Eq. \eqref{eq:Hamiltonian_matrix} are given by:
\begin{widetext}
\begin{eqnarray}
H_{x,1;x,1}(\mathbf k)&=&\epsilon_{xy} -2t_{xy} \cos\left(k_z \frac{c}{a}\right)\; ,  \nonumber \\
H_{x,1;x,2}(\mathbf k)&=&t_{\bot}+\left( \frac{3}{4}t_{\|}
         +\frac{1}{4}t_{\bot} \right)\exp\left(-i \frac{3}{2}k_y \right)
         2\cos\frac{\sqrt{3}}{2}k_x \;, \nonumber \\
H_{x,1;y,2}(\mathbf k)&=& -\frac{\sqrt{3}}{4}\left( t_{\|}-t_{\bot}\right)
         \exp\left(-i \frac{3}{2}k_y \right)
         2i\sin\frac{\sqrt{3}}{2}k_x \; ,   \label{eq:matrix} \\
H_{y,1;y,2}(\mathbf k)&=& t_{\|}+\left( \frac{3}{4}t_{\bot}
         +\frac{1}{4}t_{\|} \right)\exp\left(-i \frac{3}{2}k_y \right)
         2\cos\frac{\sqrt{3}}{2}k_x \; , \nonumber \\
H_{z,1;z,1}(\mathbf k)&=& \epsilon_{z}-2t_z  \cos\left(k_z \frac{c}{a}\right) \; , \nonumber \\
H_{z,1;z,2}(\mathbf k)&=& t\left ( 1+\exp\left(-i\left(
         \frac{\sqrt{3}}{2}k_x+\frac{3}{2}k_y  \right) \right)\right)\; . \nonumber
\end{eqnarray}
\end{widetext}
All the other matrix components of the
Hamiltonian matrix can be written in terms
of those given in Eq. \eqref{eq:matrix} as follows:
$H_{y,1;y,1}(\mathbf k)=H_{x,2;x,2}(\mathbf k)
=H_{y,2;y,2}(\mathbf k)=H_{x,1;x,1}(\mathbf k)$,
$H_{y,1;x,2}(\mathbf k)=H_{x,1;y,2}(\mathbf k)$,
$H_{z,2;z,2}(\mathbf k)=H_{z,1;z,1}(\mathbf k)$.
All the other elements are equal to zero.
This matrix is Hermitian, $H_{ji}=H_{ij}^{*}$ for $i\ne j$.
The best fit to other calculated band
structure\cite{Mazin} is obtained for the following
set of parameters: $ \epsilon_{xy}=-8.6\,{\rm eV}$,
$\epsilon_z=-1.5\,{\rm eV}$,
$t={\rm 2.0\,eV}$, $t_z={\rm 2.5\,eV}$, $t_{\|}={\rm 4.5\,eV}$,
$t_{\bot}={\rm 1.8\,eV}$, $t_{xy}={\rm 0.1\,eV}$.
In our calculation, $t$ and $t_z$ are the hopping integral
corresponding to the $p_z$ orbitals:
$t$ is the in-plane hopping between the nearest neighbors ($\pi$ bonding),
and $t_z$ is the out of plane hopping ($\sigma$ bonding). The parameters
$t_{\|}$ and $t_{\bot}$ denote
$\sigma$ and $\pi$-like hopping integrals for the in-plane 
 $p_{x,y}$ orbitals. Finally, 
the out of plane hopping integral is given by $t_{xy}$. This parameter
is very small, so 
there is practically no dispersion along the $\Gamma-A$ line.
\begin{figure}[h]
\centerline{\includegraphics[width=3.0in]{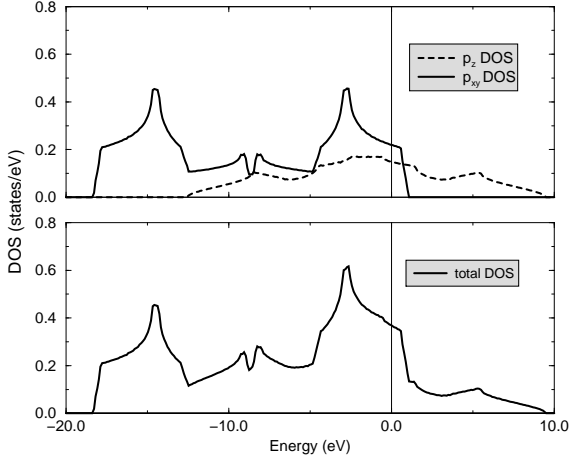}} \vspace*{3ex}
\caption{Top: Density of states for $p_{x,y}$ and $p_z$ bands. Bottom: Total
density of states. The vertical line represents the position of the Fermi energy.}
\label{fig:density}
\end{figure}
The corresponding density of states (DOS) has been calculated
in the framework of Green's function formalism as
$\varrho(\omega)=-{1 \over \pi}\sum_{\mathbf k} {\rm Im} G_b\left(\mathbf{k}, \omega \right)$
where $G_b\left(\mathbf{k}, \omega \right)$ is the Green's function corresponding to every 
band $b$.
The resulting DOS is presented in Fig. \ref{fig:density}.
The values for the DOS at the Fermi surface for the bands that cross the Fermi surface are
$\varrho_{x}=0.081\, {\rm states/eV} $, $\varrho_{y}=0.13\, {\rm states/eV} $,
$\varrho_{z1}=\varrho_{z2}=0.75\, {\rm states/eV}$, in reasonable 
agreement with more sophisticated band structure calculations\cite{Mazin}.

\section{Average over the Fermi Surface}

Throughout our analysis we have to evaluate averages over the Fermi surface.
For a given band we have to calculate
\begin{equation}
\frac{1}{S_b}\int_{S_b}\varphi\left(\mathbf k \right ) d^2 \mathbf k\; ,  \label{eq:B1}
\end{equation}
where $S_b$ represents the Fermi surface area for band $b$ and
$\varphi\left(\mathbf k \right )$ is a momentum dependent function.
The Fermi surface
was obtained in our calculation by numerically solving the equation
$\varepsilon_{\mathbf k, b }= 0$. To evaluate \eqref{eq:B1}
we replace the integration over the Fermi
surface with an integration over an energy
shell of thickness $d\varepsilon$. First
the area of the Fermi surface can be calculated as
\begin{equation}
S_b = \int_{S_b}d^2 \mathbf k=\frac{1}{d\varepsilon} \int_{shell}d^3
      \mathbf k | \nabla \varepsilon_{\mathbf k, b }|\; . 
\end{equation}
In a similar way the average of any momentum--dependent function can be
evaluated as:
\begin{equation}
\int_{S_b} \varphi\left(\mathbf k \right ) d^2 \mathbf k
              =\frac{1}{d\varepsilon} \int_{shell}d^3
               \mathbf k | \nabla \varepsilon_{\mathbf k, b }|\varphi\left(\mathbf k \right )\; .
\end{equation}
Our quantity is therefore given by the expression:
\begin{equation}
\frac{1}{S_b}\int_{S_b}
\varphi\left(\mathbf k \right ) d^2 \mathbf k =\frac{ \int_{shell}d^3
                         \mathbf k | \nabla \varepsilon_{\mathbf k, b }|\varphi\left(\mathbf k \right )}
                        { \int_{shell}d^3
      \mathbf k | \nabla \varepsilon_{\mathbf k, b }|}\label{eq:average}\; .
\end{equation}
In the numerical calculations we used a $100\times
100\times 100$ discretization of the first Brillouin zone. 
For each site in the
discretized lattice we  calculated the energy values corresponding
to band $b$. We then tested if  
one of these $10^{6}$ cells overlapped with the shell of
thickness $d\varepsilon$. If it did, we generated a mesh of $15\times 15\times
15$ within this cell  to compute the cell's contribution to Eq. \ref{eq:average}.

In our calculations, the number of points around the Fermi surface was
larger than $10^6$ within an energy shell of $10{\rm meV}$. This was
used to evaluate the average of the form factors and their Fourier
transforms  at the Fermi surface with a precision of $\sim 10^{-3}$.

\section{Momentum summation}
In this section we explain the method that we used
to evaluate the momentum summation in the
first Brillouin zone.
During the calculations we have to evaluate
expression of the form:
\begin{equation}
\frac{1}{V}\sum_{\mathbf k}
\varphi\left( \mathbf k\right)
G_b^{(0)}\left ( \mathbf k,\omega \right )\; , 
\end{equation}
where $\varphi\left( \mathbf k\right)$ is a momentum dependent
function (usually the form factor or a combination including form
factors and other momentum--dependent functions) and  $
G_b^{(0)}\left(\mathbf k,\omega \right) $ is the free Green's
function. The free Green's function depends on momentum only through
the energy of the given band $\varepsilon_{\mathbf k, b }$. We
approximated therefore the summation as
\begin{eqnarray}
&& \frac{1}{V}\sum_{\mathbf k} \varphi\left( \mathbf k\right) G_b^{(0)}\;\left ( \mathbf k,\omega \right )
 \rightarrow \\
&&\rightarrow \varrho_b \int\limits_{-\infty}^{\infty} d\varepsilon\;
                       G_b^{(0)}\left ( \varepsilon, \omega \right )
                       \frac{1}{S_b}\int_{S_b}d^2 \mathbf k
               \varphi\left( \mathbf k\right)\; , 
\end{eqnarray}
with $\varrho_b$ the density at the Fermi surface in band $b$.
The integration of the Green's function
over the energy can be done analytically and the result is simply
\begin{equation}
F_b(\omega )=\varrho_b\int\limits_{-\infty }^\infty d\varepsilon \;{\frac
1{\omega -\varepsilon \tau ^z-\Delta _b\tau ^x}}\;,
\end{equation}
For $|\omega |<\Delta _b$ the function $F_b(\omega )$ has only real parts and
simplifies to $F_b(\omega )=-\pi \varrho _b(\omega +\Delta _b\tau
^x)/(\Delta _b^2-\omega ^2)^{-1/2}$, while for $|\omega |>\Delta _b$ it is
purely imaginary.
The other term which represents an average over the Fermi surface was calculated
numerically as explained in Appendix B.


%
\end{document}